\documentclass[a4paper,10pt]{article}
\pdfoutput=1

\usepackage{graphicx}
\usepackage{hyperref}
\usepackage{amsmath}
\usepackage{bm}
\begin{document}

\title{A Partially Non-Ergodic Ground-Motion Prediction Equation for Europe and the Middle East
}

\author{Nicolas M.\ Kuehn\\
Pacific Earthquake Engineering Research Center\\
University of California, Berkeley
\and
Frank Scherbaum\\
Insitute of Earth and Environmental Science\\
University of Potsdam
}

\maketitle

\begin{abstract}
A partially non-ergodic ground-motion prediction equation is estimated for Europe and the Middle East.
Therefore, a hierarchical model is presented that accounts for regional differences.
For this purpose, the scaling of ground-motion intensity measures is assumed to be similar, but not identical in different regions.
This is achieved by assuming a hierarchical model, where some coefficients are treated as random variables which are sampled from an underlying global distribution.
The coefficients are estimated by Bayesian inference.
This allows one to estimate the epistemic uncertainty in the coefficients, and consequently in model predictions, in a rigorous way.
The model is estimated based on peak ground acceleration data from nine different European/Middle Eastern regions.
There are large differences in the amount of earthquakes and records in the different regions.
However, due to the hierarchical nature of the model, regions with only few data points borrow strength from other regions with more data.
This makes it possible to estimate a separate set of coefficients for all regions.
Different regionalized models are compared, for which different coefficients are assumed to be regionally dependent.
Results show that regionalizing the coefficients for magnitude and distance scaling leads to better performance of the models.
The models for all regions are physically sound, even if only very few earthquakes comprise one region. 

\end{abstract}

\section{Introduction}

An important question in engineering seismology is whether ground-motion predictions equations (GMPEs) are regionally dependent \cite{Douglas2007}.
For example, in Europe both regional GMPEs (e.g.\ \cite{Akkar2010a} for Turkey; \cite{Bindi2011} for Italy; \cite{Bragato2005} for the Eastern Alps; \cite{Danciu2007} for Greece) as well as pan-European GMPEs (e.g.\ \cite{Ambraseys2005,Akkar2010}) have been developed.
The recent pan-European GMPEs described in \cite{Douglas2014}, \cite{Akkar2014,Bora2014,Derras2014,Hermkes2014} do not include regional differences.
From the models developed recent NGA-West 2 project \cite{Bozorgnia2014}, four \cite{Abrahamson2014,Boore2014,Campbell2014,Chiou2014} include regional adjustment terms.
Generally, these models start out from a global model, which is then regionally adjusted based on residual analysis.
For California, differences between Northern and Southern California have been found \cite{Atkinson2009,Chiou2010}.

Incorporating regional differences in ground-motion scaling into PSHA is a first step to remove the ergodic assumption from seismic hazard estimation \cite{Anderson1999,Stafford2014,Kuehn2015a}.
Under the ergodic assumption states the distribution of ground motions at a given site is the same as the spatial distribution over all sites given the same magnitude, distance and site conditions \cite{Anderson1999}.
Regionally dependent GMPEs relax this assumption by breaking a data set from a large region into smaller units.

Generally, it makes sense to assume that there are regional differences in ground-motion scaling between different regions, as geological conditions differ.
However, one problem that one faces when developing regional GMPEs is lack of data for smaller regions.
Recently, \cite{Gianniotis2014} showed a way to constrain coefficients to be similar across regions, and also provided evidence of better predictive performance as opposed to a global model.
In a similar vein, \cite{Stafford2014} shows how regional difference between countries (regions) can be included into a GMPE.

Here, we investigate differences in GMPEs between regions in Europe and the Middle East, based on data from the RESORCE data base \cite{Akkar2014a}.
We develop a multi-level model for peak ground acceleration (PGA) that accounts for regional differences from the start.
Therefore, the coefficients of the model are regionalized but assumed to be sampled from a global distribution -- this allows the different regions to borrow statistical strength from each other.
We investigate the predictive performance of several models, which differ in the amount of coefficients that are regionalized.
The model presented in this work shares similarities with the work of \cite{Gianniotis2014,Kuehn2015,Stafford2014}. We use the same data as \cite{Gianniotis2014}, but constrain the coefficients for different regions in a different way.
The model is cast in the formulation of \cite{Kuehn2015}. The integration of regionalized coefficients is similar to \cite{Stafford2014}, but the model presented here is estimated on a different data set. In addition, we show the effect of different regionalized models on median predictions, and how predictive performance can be used for model selection.

In general, regionalization of coefficients improves predictive performance and decreases the associated aleatory variability (see section \emph{Results}), which has consequences for seismic hazard calculations \cite{Bommer2006}.

\section{Data}

\begin{figure}
  \begin{center}
    \includegraphics[width = 0.8\textwidth]{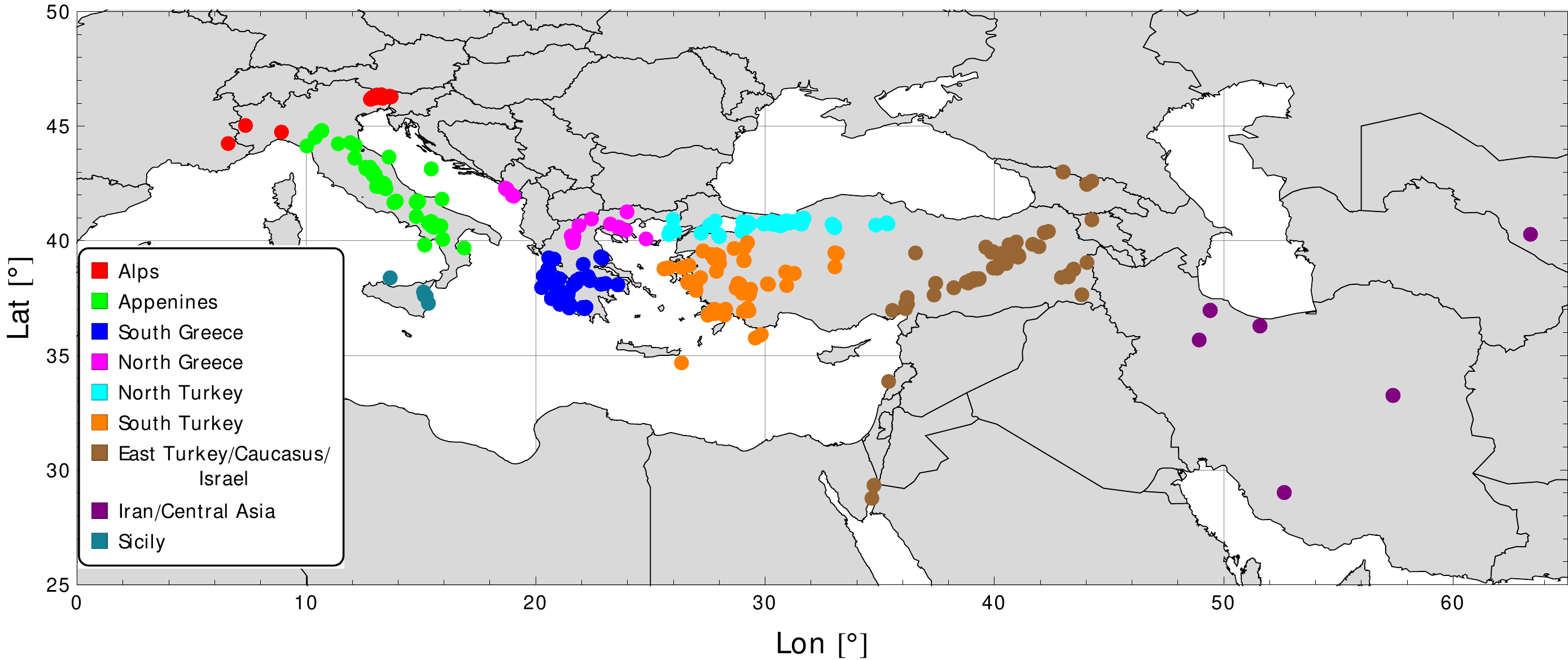}
  \end{center}
  \caption{Events in the data set, color-coded by region.}
  \label{fig: map}
\end{figure}

\begin{figure}
  \begin{center}
    \includegraphics[width = 0.95\textwidth]{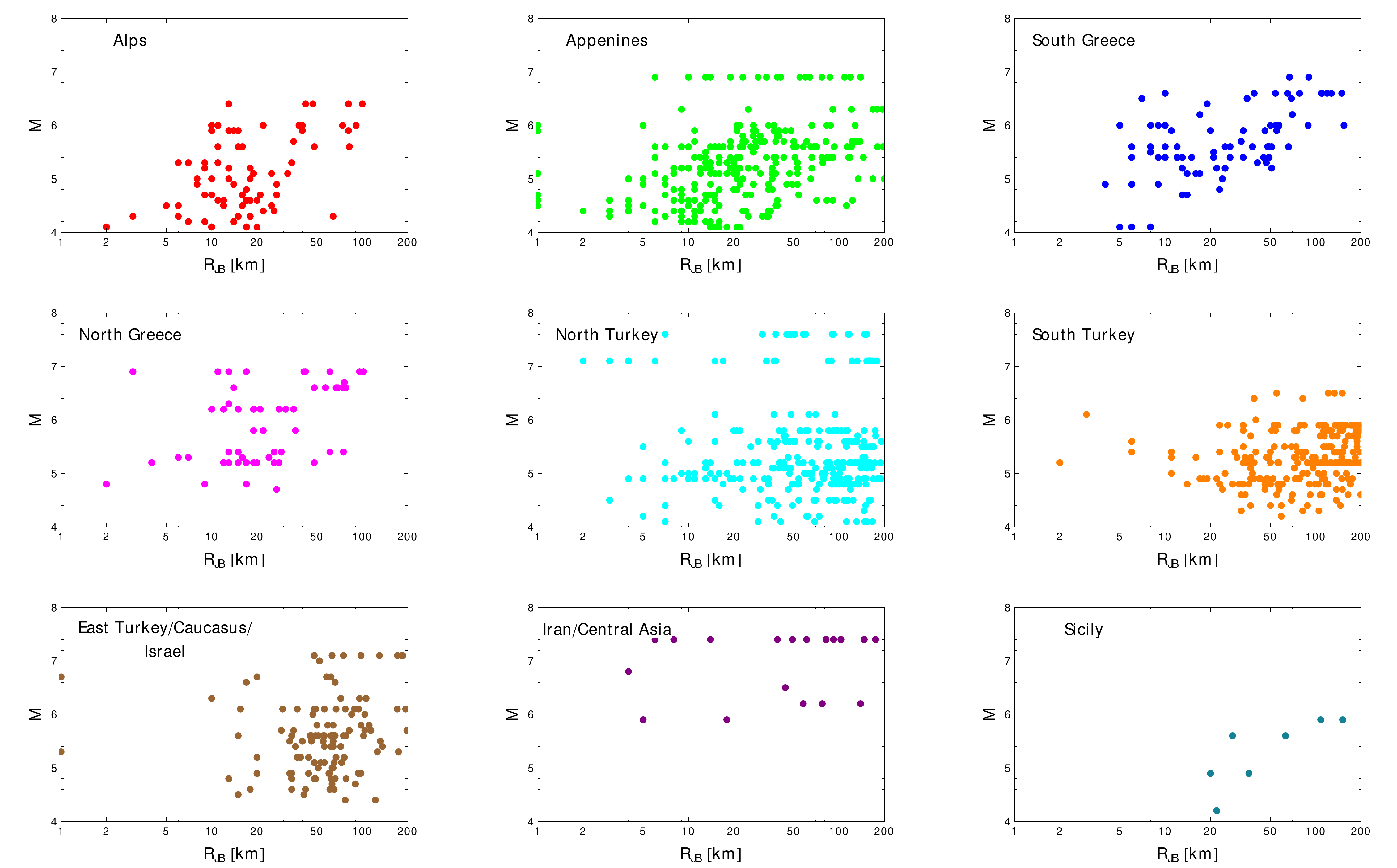}
  \end{center}
  \caption{Magnitude-distance distribution. Colors are the same as in Figure \ref{fig: map}.}
  \label{fig: MR}
\end{figure}

We use the same data set that was used by \cite{Gianniotis2014}.
It is a subset of the RESORCE data base \cite{Akkar2014a}, which was compiled for the SIGMA project\footnote{\url{http://projet-sigma.com/}}.
We provide only a brief description here.
More information can be found in \cite{Gianniotis2014} and \cite{Akkar2014a}.
In total, there are 1261 records from 362 earthquakes, recorded at 359 stations.
The events are divided into 9 regions, mainly based on geographical considerations, but informed by tectonic units defined for the SHARE project (Seismic Hazard Harmonization in Europe)\footnote{\url{http://www.share-eu.org/}} \cite{Delavaud2012TowardEurope}.
The events and their regions are shown in Figure \ref{fig: map}.
The magnitude/distance distribution of the data is shown in Figure \ref{fig: MR}.
Table \ref{tab: data} summarizes the data for the different regions.
As one can see, some regions have only a very limited magnitude/distance range.
However, as shown later, we can still estimate a physically plausible model for those regions.

\begin{table}
  \caption{number of data for the different regions}
  \begin{tabular}{c|cccc}
    Region & $N_{eq}$ & $N_{rec}$ & $M$-range & $R_{JB}$-range\\
    \hline
    Alps & 29 & 91 & 4.1 -- 6.4 & 2. -- 100. \\
    Appenines & 78 & 303 & 4.1 -- 6.9 & 0. -- 200. \\
    South Greece & 41 & 87 & 4.1 -- 6.9 & 0. -- 154. \\
    North Greece & 18 & 58 & 4.7 -- 6.9 & 2. -- 102. \\
    North Turkey & 53 & 324 & 4.1 -- 7.6 & 0. -- 191. \\
    South Turkey & 70 & 250 & 4.2 -- 6.5 & 0. -- 199. \\
    East Turkey & 62 & 122 & 4.4 -- 7.1 & 0. -- 198. \\
    Iran & 6 & 19 & 5.9 -- 7.4 & 4. -- 175. \\
    Sicily & 5 & 7 & 4.2 -- 5.9 & 20. -- 151.
  \end{tabular}
  \label{tab: data}
\end{table}

\section{Model Formulation}\label{sec: model}

We use a slightly adjusted multi-level formulation of \cite{Kuehn2015} to formulate the model:
\begin{align}
  Y_{es} &\sim N(f(\bm{x}) + \eta_e + \lambda_s,\phi_{SS}) \label{eq: Y} \\
  \eta_e &\sim N(0,\tau) \label{eq: event term} \\
  \lambda_s &\sim N(0,\phi_{S2S}) \label{eq: station term}
\end{align}
Equation \eqref{eq: Y} reads as: the ground-motion value $Y$ from event $e$ recorded at station $s$ is distributed according to a normal distribution with a mean that is a function of the predictors $\bm{x}$ plus the event term $\eta_e$ and the station term $\lambda_s$, and standard deviation $\phi_{SS}$, which denotes single-station within-event variability.
The event and station terms are distributed normally with standard deviation $\tau$ and $\phi_{S2S}$, which describe between-event and station-to-station variability \cite{Al-Atik2010}.
The ground-motion parameter of interest $Y$ is logarithmic PGA.

We use a slightly different model than \cite{Gianniotis2014}.
The predictor variables are the moment magnitude $M$, the Joyner-Boore distance $R_{JB}$, the shear wave velocity averaged over the top 30m $V_{S30}$ and the focal meachanism.
The dependence of the median predictions on the predictors is modeled by he following functional form:
\begin{align}
  f(\bm{x}) &= c_{0} + c_{1} M_e+ c_{2} M_e^2 + c_{3} F_{R} + c_{4} F_{N} + (c_{5} + c_{6} M_e) \ln \sqrt{R_{es}^2 + c_{7}^2} \nonumber \\ 
  &\quad + c_{8} R_{es} + c_{9} \ln \frac{V_{S30,s}}{760} \label{eq: functional form}
\end{align}
where $F_R$ and $F_N$ are dummy variables taking the value $1$ for reverse and normal faulting events, respectively, and $0$ otherwise.
The coefficients $c_i$ vary by region, which is indicated by index $r$.
However, to allow data from one region influence the ground-motion scaling in other regions, the coefficients are connected
\begin{align}
  c_{i} &\sim N(\mu_{c_i},\sigma_{c_i}) \label{eq: region coefficients}
\end{align}
Hence, the regional coefficients are sampled from a (global) distribution.
In terms of a multi-level model \cite{Gelman2006,Kuehn2015} the regional coefficients would be another level, a hierarchy higher than the event/station level.
In terms of a mixed-effects model, the global coefficients $\mu_{c_i}$ would be called the fixed effects, while the difference $c_{i} - \mu_{c_i}$ are the regional random-effects.
We prefer the multi-level formulation, because we believe that it is easier to understand.

The parameters of the model comprise the coefficients $c_{i,r}$, the mean and standard deviations of their global distribution, $\mu_{c_i}$ and $\sigma_{c_i}$, the components of the aleatory variability $\phi_{SS}$, $\phi_{S2S}$ and $\tau$, as well as the event and station terms $\eta_e$ and $\lambda_s$. 

The parameters of the model are estimated via Bayesian inference using the program Stan \cite{Team2015}, which performs Hamiltonian Monte Carlo sampling.
Bayesian inference relies on Bayes theorem to estimate the posterior distribution of the parameters, which is proportional to he product of the likelihood, determined by the model of equation \eqref{eq: functional form}, times the prior distribution of the parameters.
The prior distribution describes the uncertainty of the parameters before data is taken into account.
Hence, the posterior distribution describes the uncertainty of the model parameters given the observed data.
For a more detailed explanation of Bayesian inference see e.g.\ \cite{Spiegelhalter2009}.
For the parameters describing standard deviations we use Half-Cauchy distributions as prior distributions \cite{Gelman2006PriorModels,Team2015a}.
The priors for the global parameters $\mu_{c_i}$ are normal distributions, based on the model of \cite{Abrahamson2014}.
This model has been shown to extrapolate well to large magnitudes and short distances, so using it as a basis for setting the prior distributions makes these more informative than just setting wide normal distributions.
The model of \cite{Abrahamson2014} is chosen because it is predominantly based on a global data set and thus has only partial overlap with RESORCE database. Hence, data is not used twice in both the prior and the likelihood.

We evaluate the model of \cite{Abrahamson2014} at different settings of the predictor variables ($M = 4.5,5,\ldots,8$;  $R_{JB}=1,5,10,20,30,50,75,100,150,200$;  $V_{S30}=300,400,\ldots,1100$; $F_N = F_R =0,1$). Values for the depth of the top of the rupture are calculated from the model presented in \cite{Chiou2014}.
All evaluations are carried out on the footwall.
We then fit equation \eqref{eq: functional form} to to the predictions of the model of \cite{Abrahamson2014}.
The fitted coefficients provide the mean values for the prior distributions.
Ideally, the width of the prior distributions would be set by taking into account the uncertainty of the model of \cite{Abrahamson2014}, together with the uncertainty of the fitted coefficients.
Since the former uncertainty is not readily available, we fit equation \eqref{eq: functional form} to the data and use the standard error of the coefficients, together with the standard error of the fit to the model of \cite{Abrahamson2014}, for the standard deviations of the prior distributions for the coefficients.

To avoid unphysical behavior of the expected models, we constrain the parameters $c_{i}$ and $\mu_{c_i}$ to be either positive or negative, except for the constant term $c_0$.

The parameters are estimated by sampling from the posterior distribution using the program Stan \cite{Team2015}.
We run 4 chains of samples, with different starting values.
To avoid influence of the starting values, we discard a burn-in of the first 1000 samples.
After the burn-in period, each chain is run for another 1000 samples. 
We keep every fifth sample, leading to 800 draws from the posterior distribution.
Convergence of the chains is assessed using the Gelman-Rubin statistic \cite{Gelman1992}.

\section{Model Selection}

In the previous section we have formulated a multi-level ground-motion model that can take regional differences into account by constraining the parameters via equation \eqref{eq: region coefficients}.
This can be done for all coefficients, or just for a subset.
It is also possible to estimate a model that pools all data together into one region.
Furthermore, we can estimate a separate model for each region, where the model coefficients are not connected via equation \eqref{eq: region coefficients}.
It is important to test whether a regionalized model performs better than a “global” model or individual models.
One often used measure to compare different models is the Akaike Information Criterion (AIC) \cite{Akaike1973}.
However, AIC cannot be easily generalized to Bayesian models. Furthermore, the calculation of AIC requires the number of parameters in the model, which is not straightforward for multi-level models, since shared parameters (such as an event term) cannot be treated like a fixed coefficient in the model (see \cite{Gelman2013} for a discussion on these issues).

\cite{Gianniotis2014} performed experiments splitting the data set into training and test data and found that a regionalized, connected model performs generally better on the test data than the pooled model.
This already provides a compelling argument for splitting a data set into subregions, for which individual models can be estimated.

Splitting the data set into a training and test set and evaluating the model on the test set provides a measure how well the model can predict new, future data.
This is called the generalization error.
There are different possibilities how the data set could be split into training/test data.
One could select the test data such that it contains only records from events that are not in the training data.
Another way would be to have records from one event in both the training and test data set.
In the latter case, one can make use of an estimated event term from the training data, and hence get better residuals (in terms of a smaller overall residual).
Ultimately, the two strategies answer two different questions -- how well does the model predict data from a new event, and how well does it perform on an event from which data already exists.
Both of these measures are important.

It is computationally very demanding to repeatedly fit different models to training and test data.
Therefore, in this work estimate the generalization error using a measure that is based on draws from the posterior distribution of the parameters.
This measure is the widely applicable Information Criterion (WAIC) \cite{Vehtari2014,Watanabe2010AsymptoticTheory}.
The estimated quantity is $\widehat{elpd}_{WAIC}$, which is an approximation to the expected log pointwise predictive density for a new data set.
The predictive density describes the full uncertainty distribution of a prediction from a model, taking into account the uncertainty of the parameters (characterized by the posterior distribution) and the aleatory variability.
For details on theory, derivation and implementation of $\widehat{elpd}_{WAIC}$, see \cite{Vehtari2014,Vehtari2015,Gelman2013}.

The measure $\widehat{elpd}_{WAIC}$ is calculated based on the log-likelihood of the data given draws from the posterior distribution of the parameters $\theta$.
The log-likelihood can be calculated in different ways, based on the question one might want to ask.
Here, we calculate two different likelihoods, and consequently get two different values for $\widehat{elpd}_{WAIC}$:
\begin{align}
  p(y_{es}|\theta^S) &= N(f(\bm{x}) + \eta_e + \lambda_s,\phi_{SS}) \label{eq: waic1} \\
  p(y_{es}|\theta^S) &= N(f(\bm{x}),\sigma_{T}) \label{eq: waic2}
\end{align}
where $\theta^S$ is a draw from the posterior distribution of the parameters, and $\sigma_T^2 = \tau^2 + \phi_{SS}^2 + \phi_{S2S}^2$ is the total variability.

The $\widehat{elpd}_{WAIC}$ based on the likelihood calculated using equation \eqref{eq: waic1} is a measure of how well the model can predict a new record, given knowledge of the event term $\eta$ and station term $\lambda$.
On the other hand, the second likelihood describes how well the model can predict a new record, given only the median prediction $f(\bm{x})$.
Hence, equation \eqref{eq: waic1} measures how well a model can predict a new record for an already observed event, while equation \eqref{eq: waic2} provides a measure how well a model can predict a new record from an unknown event.

We use WAIC (or more precisely $\widehat{elpd}_{WAIC}$) to compare different models:
\begin{itemize}
  \item A model estimated on the whole, pooled data set, which we term ``global''.
  \item A model with independent, separate coefficients for each region, which we call ``individual'' (I).
  \item A model with coefficients that are connected via equation \eqref{eq: region coefficients}. The model where all coefficients vary by region is called ``regional 1'' (R1).
  \item A model with regionalized coefficients, where the near-source saturation term $c_6$ does not vary by region. This model is called ``regional 2'' (R2).
  \item A model where coefficients $c_3, c_4, c_7$ do not vary by region, called ``regional 3'' (R3).
  \item A model where only coefficients $c_0,c_8,c_9$ vary by region, called ``regional 4'' (R4).
  \item A model where only coefficient $c_0$ varies by region, called ``regional 5'' (R5).
\end{itemize}
In all cases, we assume the standard deviations to be the same for all regions.
This avoids trade-offs between differences in the coefficients and the standard deviations.
Implicitly, this means that the variability of stress drop and geological conditions within the regions is assumed to be similar.

\section{Results}

\begin{figure}
  \begin{center}
    \includegraphics[width = 0.45\textwidth]{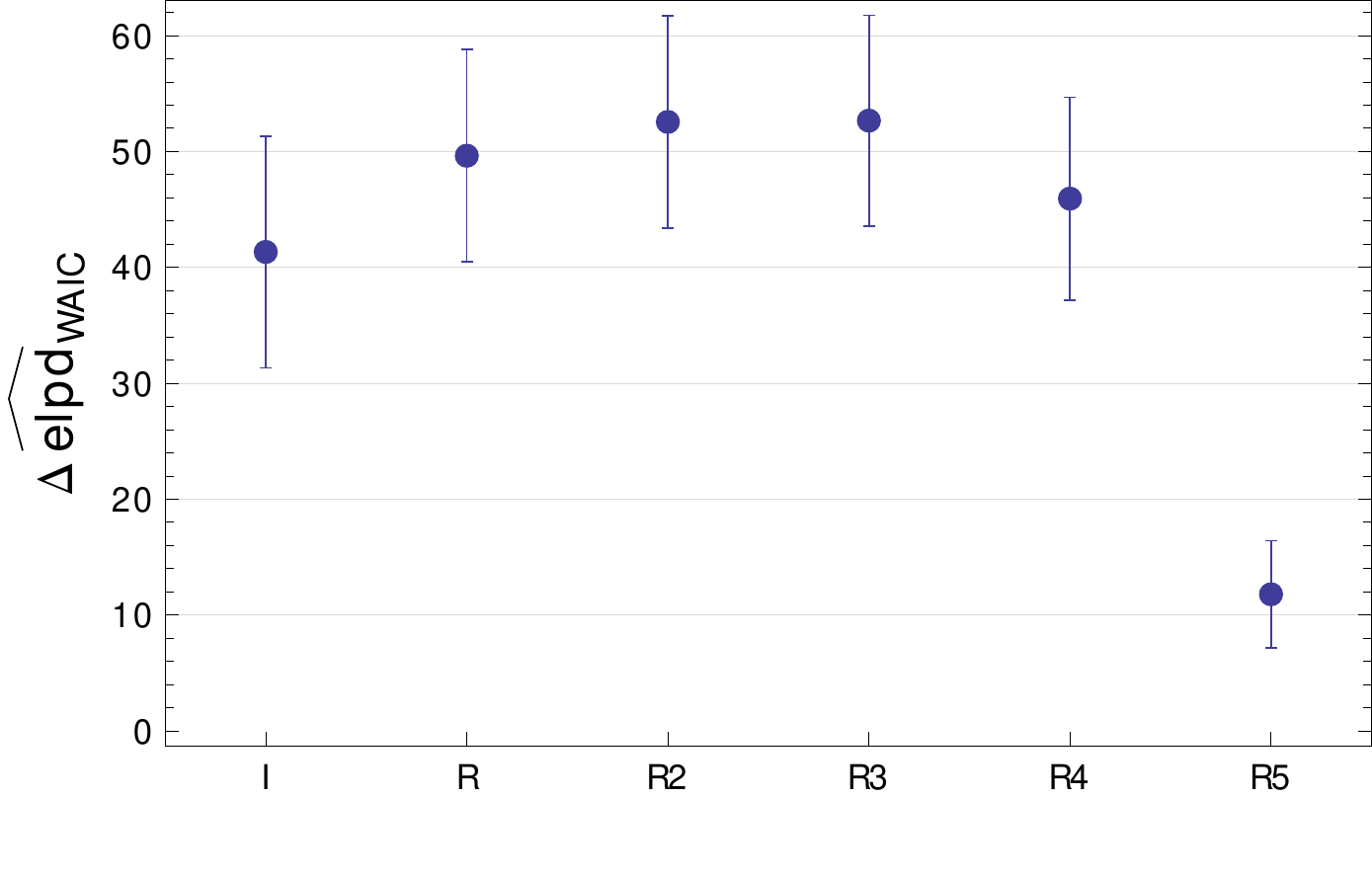}
    \hspace{0.05\textwidth}
    \includegraphics[width = 0.45\textwidth]{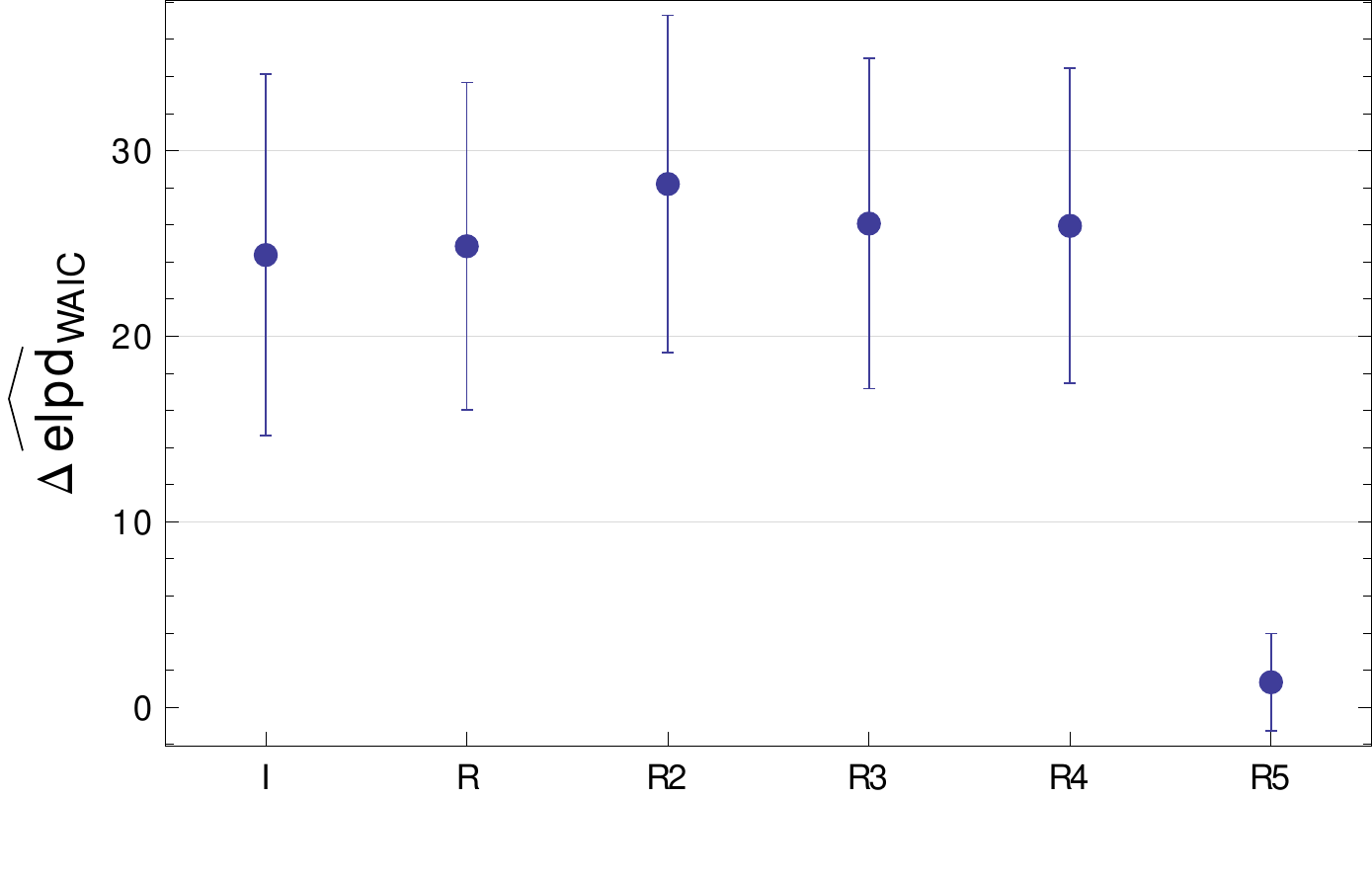}
  \end{center}
  \caption{\textbf{Left:} $\Delta\widehat{elpd}_{waic}$ and their standard error for the different models, computed based on equation \eqref{eq: waic1}; \textbf{Right:} $\Delta\widehat{elpd}_{waic}$ and their standard error for the different models, computed based on equation \eqref{eq: waic2}.}
  \label{fig: waic}
\end{figure}

Figure \ref{fig: waic} shows the difference in $\widehat{elpd}_{WAIC}$ between the regionalized models and the global model, where a positive difference indicates a better fit of the regional model.

We can see a positive $\Delta\widehat{elpd}_{waic}$ for all the regionalized models compared to the global model.
There is not a big difference between the different regionalized models, except for the ``regional 5'' model, which is close to the global model.
This indicates that there are differences in scaling with the predictor variables between the different regions, and that these differences are important for predicting new data points.
The separate model also has better predictive capability (according to $\widehat{elpd}_{WAIC}$) than the global model, and is comparable to the regional models when the event and station term are included (equation \eqref{eq: waic1}).
This is not really surprising, since knowing these terms provides a lot of information in the case of a small data set.
Furthermore, the informative prior provides some constraint on the parameters, so that even with a small number of data the estimated coefficients are not completely physically unreasonable.

\begin{figure}
  \begin{center}
    \includegraphics[width = 0.45\textwidth]{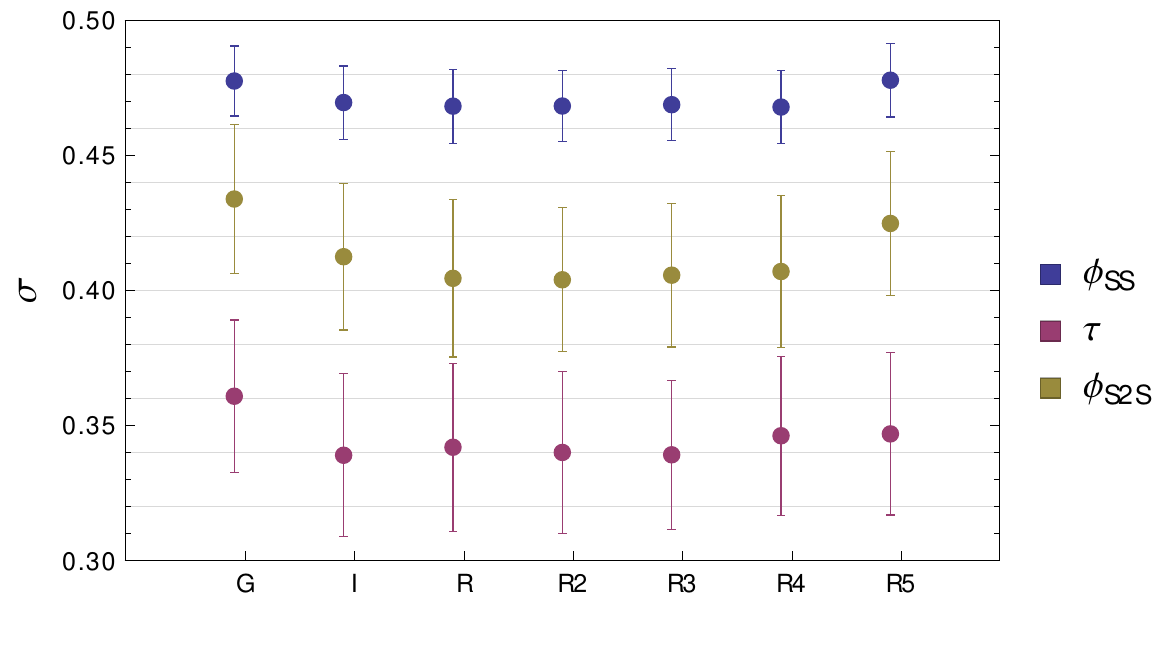}
  \end{center}
  \caption{Between-event, within-event and between-station standard deviations. Error bars show the standard deviation of the posterior distribution.}
  \label{fig: sigma}
\end{figure}

Figure \ref{fig: sigma} shows the estimated between-event, within-event and station-to-station standard deviations.
The values decrease for the regional models, with the main contribution to the lower total standard deviation coming from $\phi_{S2S}$.
This indicates that there are mainly differences in site scaling between the different regions.

The values of the $\widehat{elpd}_{WAIC}$ parameters provide some indication that a hierarchical, regionalized model might be beneficial in terms of ground-motion modeling.
However, $\widehat{elpd}_{WAIC}$ is an approximation to the expected log pointwise predictive density for a new data set, where the new data set may include new events and stations, but does not go beyond the ranges of the predictor variables.
For an application of a ground-motion model in PSHA, however, it is important that the model extrapolates, e.g.\ to large magnitudes and short distances, in a physically meaningful way.
In particular, some of the regions have a pretty limited coverage of magnitudes and distances (cf.\ Figure \ref{fig: MR}), so it is important to ensure that they extrapolate well to $M/R$-ranges of interest.
On the one hand, using an informative prior on the parameters (based on the model of \cite{Abrahamson2014}) constraints the variability of the parameters and should lead to reasonable behavior.
However, since the coefficients for the individual regions are only indirectly constrained by the prior via equation \eqref{eq: region coefficients}, it is worthwhile to check their scaling.
In the following, we use the ``regional 3'' model, for which all coefficients except $c_3$, $c_4$ and $c_7$ vary by region.
According to Figure \ref{fig: waic} is has good generalization performance.
It makes sense to not regionalize the scaling with focal mechanism since the distribution of focal mechanisms across regions is very uneven.

\begin{figure}
  \begin{center}
    \includegraphics[width = 0.45\textwidth]{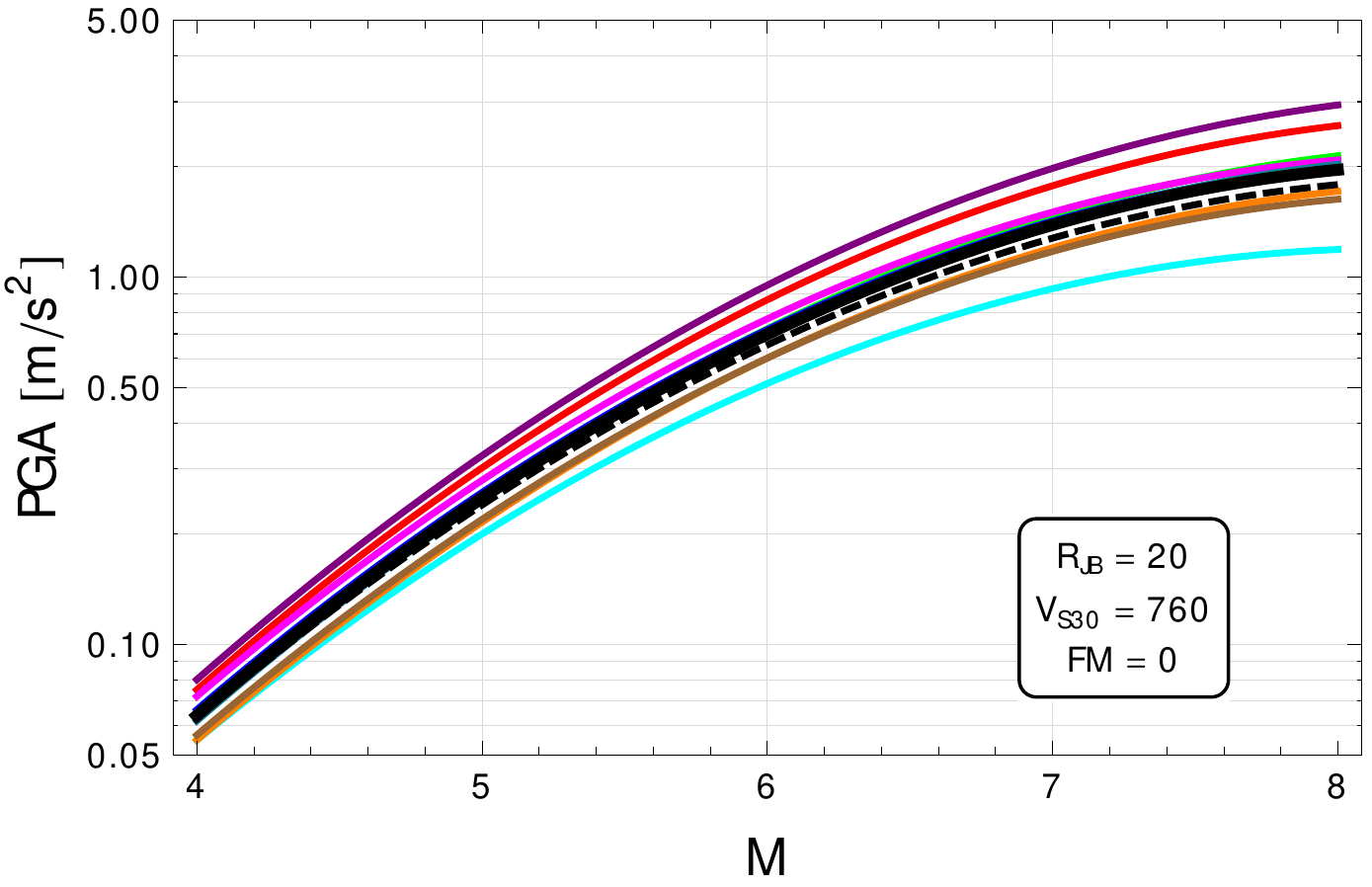}
    \hspace{0.05\textwidth}
    \includegraphics[width = 0.45\textwidth]{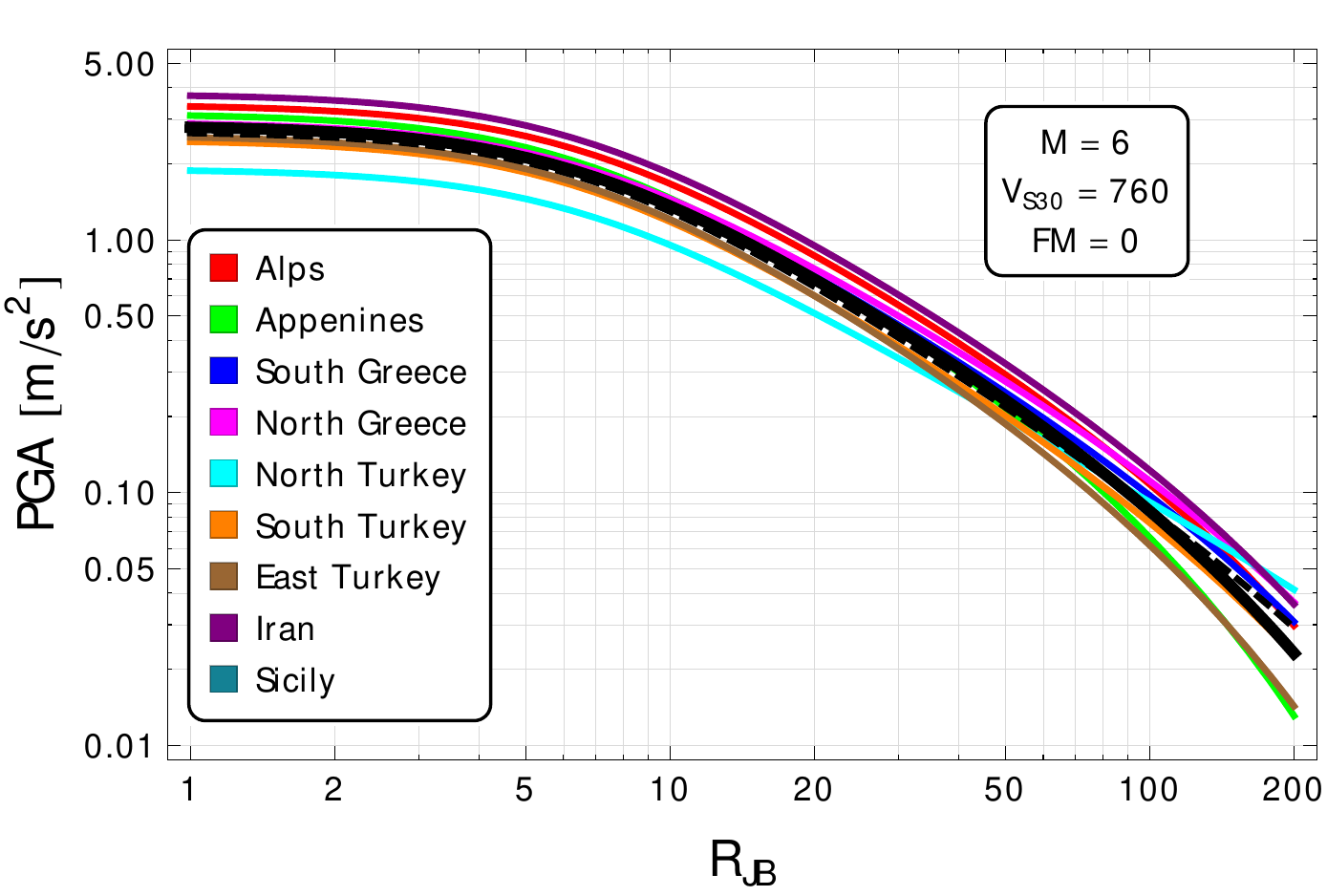}
  \end{center}
  \caption{Scaling of regional models with magnitude (\textbf{left}) and distance (\textbf{right}). The thick black line is the scaling using the global parameters $\mu_{c_i}$. The dashed black line is a model based on no regionalization. Colors are the same as in Figure \ref{fig: map}.}
  \label{fig: scaling}
\end{figure}

In Figure \ref{fig: scaling}, the scaling of PGA with magnitude and distance is shown, using the regionalized coefficients.
The curves are based on the mean of the posterior distributions for the individual parameters.
As one can see, the models for all regions have a physically plausible shape -- even for regions such as Sicily or Iran, which have a very limited number of indigenous data.
In these cases, their model borrows strength from the other regions via the hierarchical model formulation of equation \eqref{eq: region coefficients}.
One can see in Figure \ref{fig: scaling} that the regional models for Turkey produce relatively low median predictions.
This is concordant with the models of \cite{Akkar2010a,Bora2014}, which both produce relatively low ground-motion predictions and which contain a large percentage of Turkish data.

\begin{figure}
  \begin{center}
    \includegraphics[width = 0.3\textwidth]{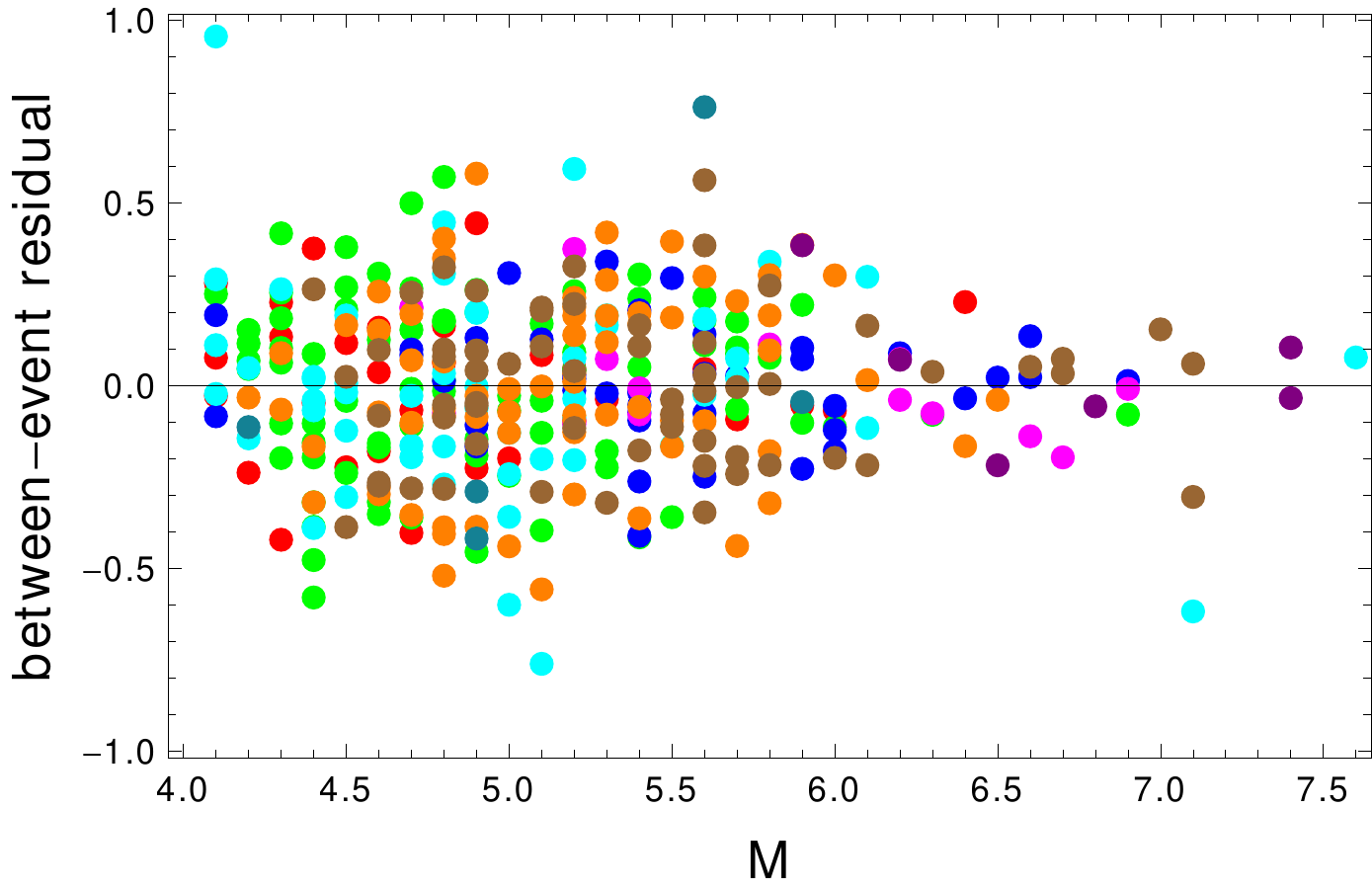}
    \hspace{0.02\textwidth}
    \includegraphics[width = 0.3\textwidth]{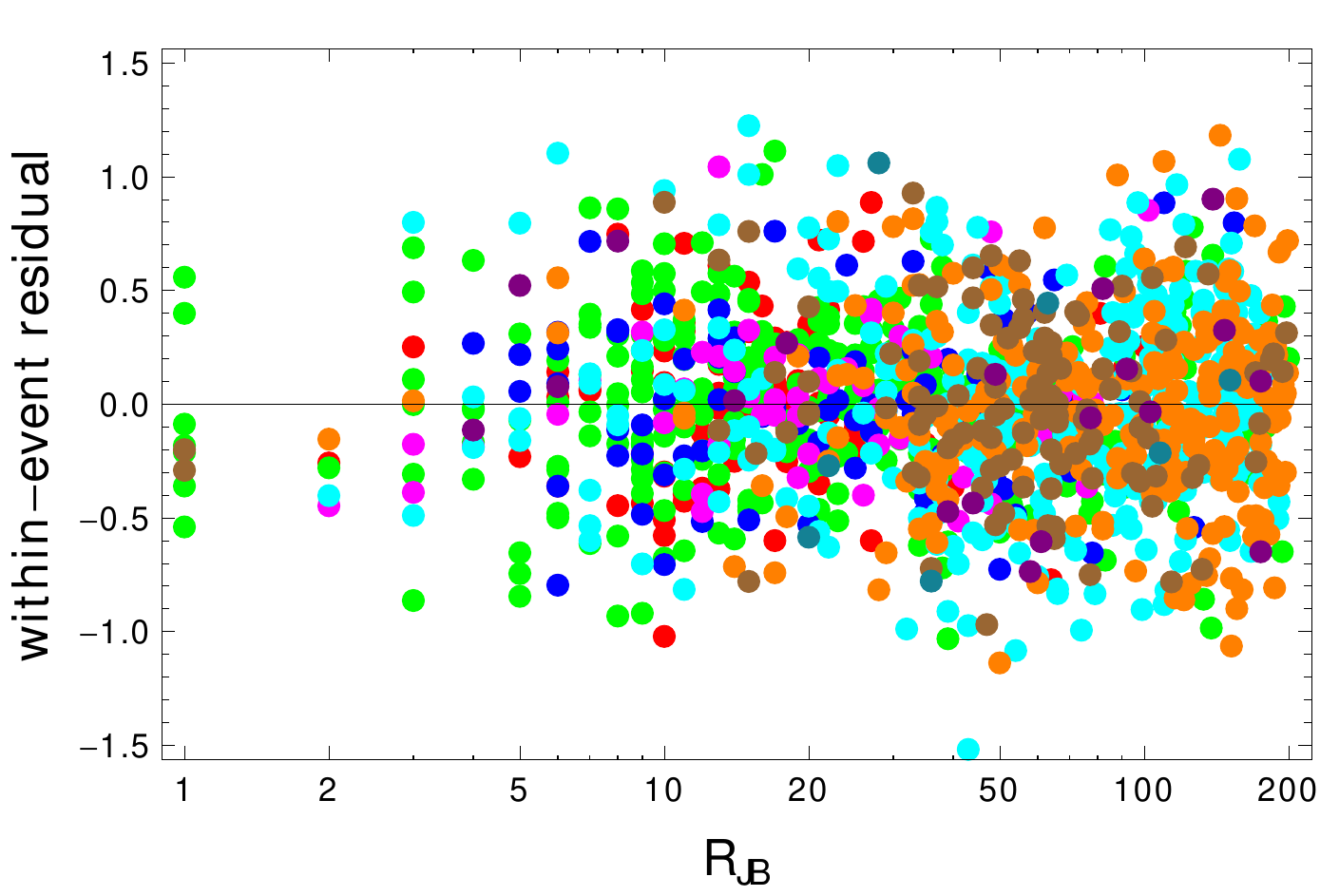}
    \hspace{0.02\textwidth}
    \includegraphics[width = 0.3\textwidth]{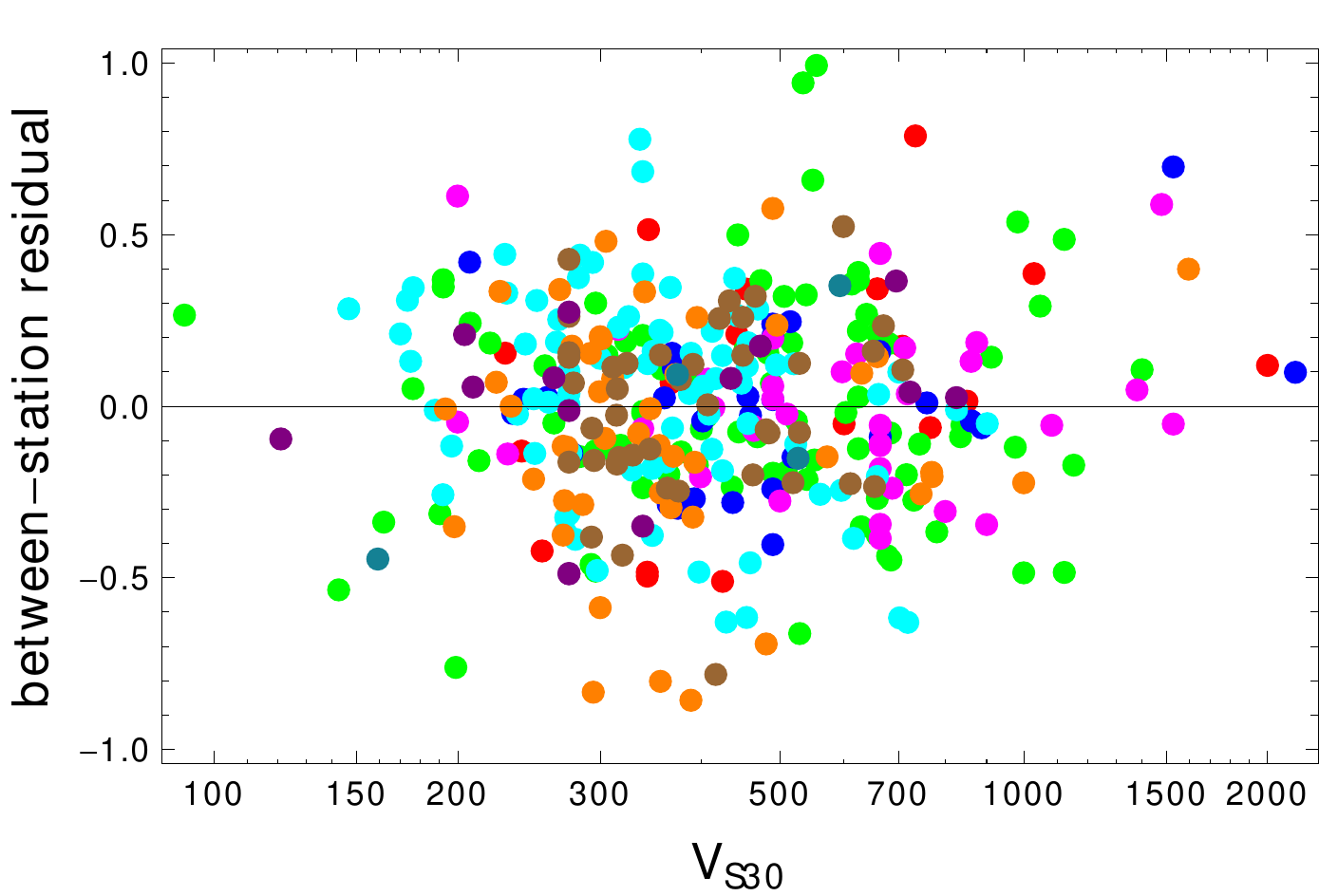}
  \end{center}
  \caption{Between-event, within-event and station-to-station residuals, colorcoded by region.}
  \label{fig: residuals}
\end{figure}

\begin{figure}
  \begin{center}
    \includegraphics[width = 0.45\textwidth]{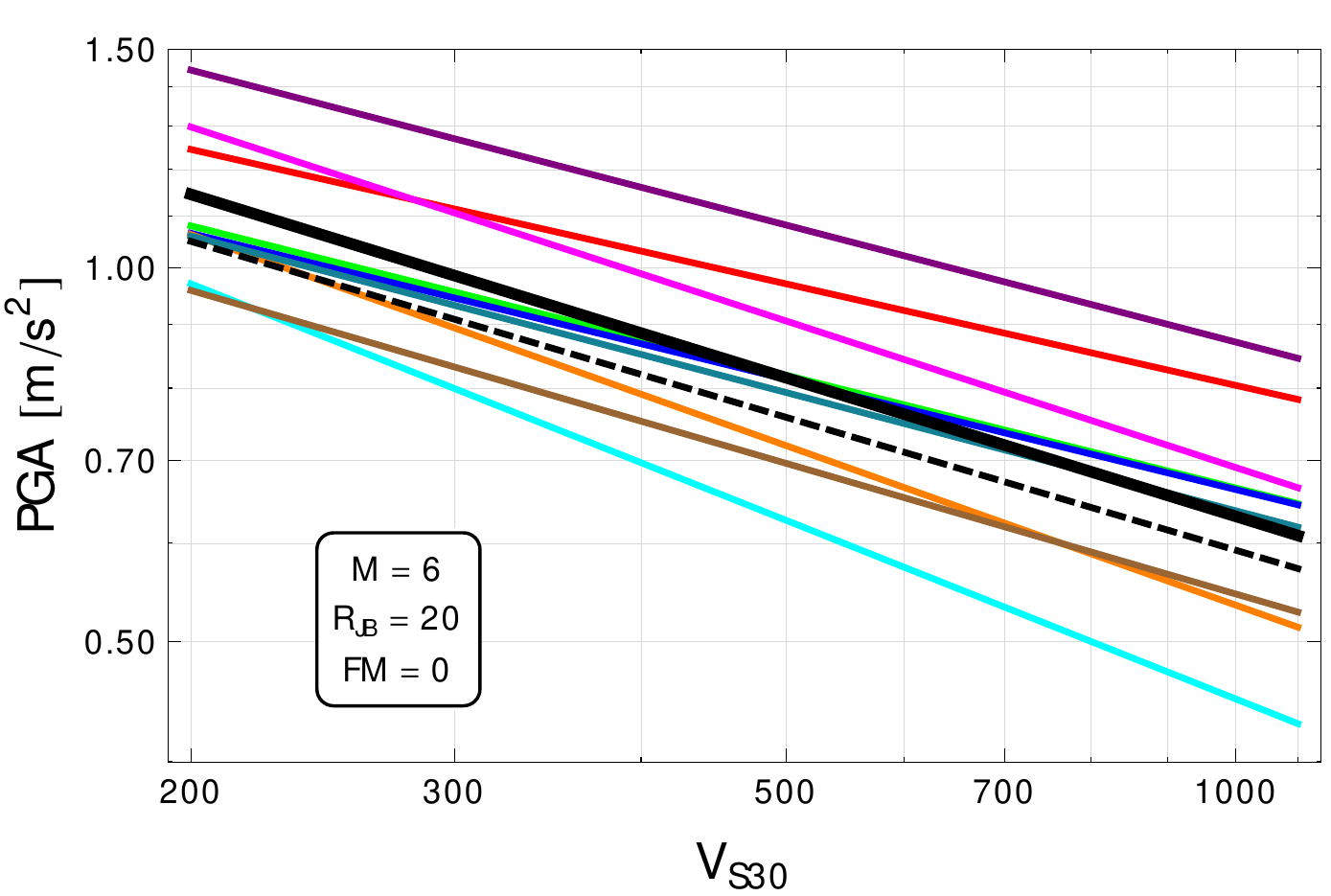}
  \end{center}
  \caption{Scaling of regional models with $V_{S30}$.}
  \label{fig: scaling vs}
\end{figure}

The residuals of the model are shown in Figure \ref{fig: residuals}.
They are unbiased for all regions, indicating that the model formulation works well in fitting the data, and also in accounting for regional differences.
Since the main reduction in the standard deviations is for the site-to-site variability $\phi_{S2S}$, the scaling of PGA with $V_{S30}$ is shown in Figure \ref{fig: scaling vs}.
The slopes are quite different for some regions, which explains the lower standard deviation.

\section{Discussion and Conclusions}

We have shown a way to estimate regionalized GMPEs, using a multi-level approach, as previously explored by \cite{Stafford2014}.
Using this approach, it is possible to estimate individual, physically sound models even for regions that do not comprise a lot of data, by constraining their coefficients to be close to the global coefficients.
The amount by which they can differ is determined by the data.

We have seen in Figure \ref{fig: waic}, that the hierarchical regionalized model gives higher values of $\widehat{elpd}_{WAIC}$ than a global model that pools all data, which means that the regionalized model has higher generalization capability.
Furthermore, the models for the individual regions are physically plausible.

\begin{figure}
  \begin{center}
    \includegraphics[width = 0.45\textwidth]{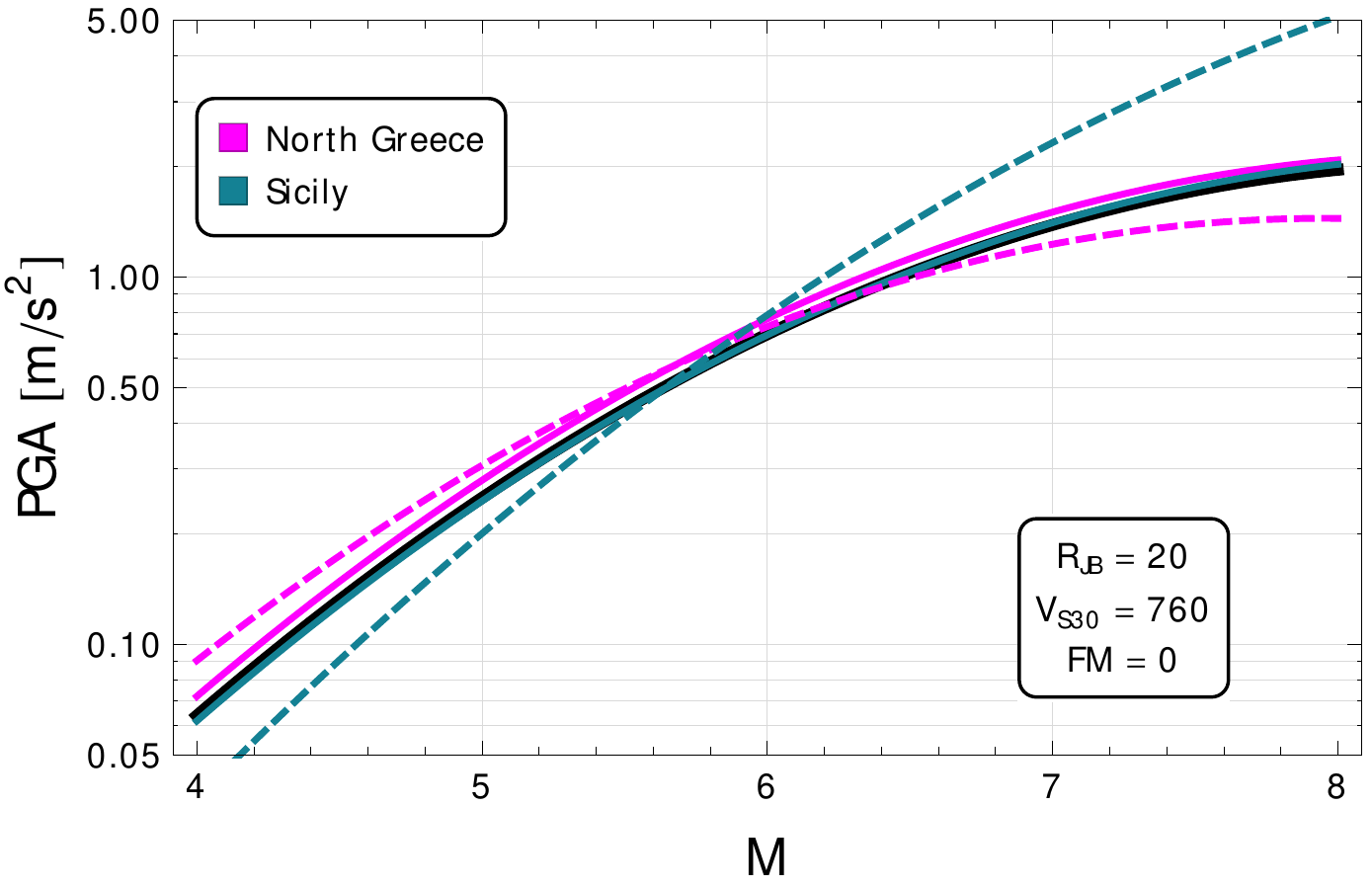}
  \end{center}
  \caption{Scaling of regionalized models for two regions, calculated using the ``regional 3'' model (solid lines) and a model with independent coefficients (``individual'', dashed lines). Black line is the global model. The predictions of the regionalized models are drawn towards the global model in the case of only few data points in the regions.}
  \label{fig: comparison}
\end{figure}

In Figure \ref{fig: comparison}, we illustrate the effect of the constraint imposed by equation \eqref{eq: region coefficients}.
The magnitude scaling is shown for Sicily and Northern Greece, which both have relatively few data points, for the regionalized model (R3) and the ``individual'' model.
We can see that in the regionalized model, the curves are pulled towards the global model -- the models for these regions are constrained by data from the other regions.
This effect makes it possible to extrapolate the regionalized models to data ranges (for example higher magnitudes or shorter distances) that are not covered by data for the individual regions – in these ranges, the fact the regional coefficients are conencted via equation \eqref{eq: region coefficients} ensures that the regional models behave similarly to the global model.
Hence, we believe that the proposed methodology is a good way to estimate regional ground-motion models.

\begin{figure}
  \begin{center}
    \includegraphics[width = 0.45\textwidth]{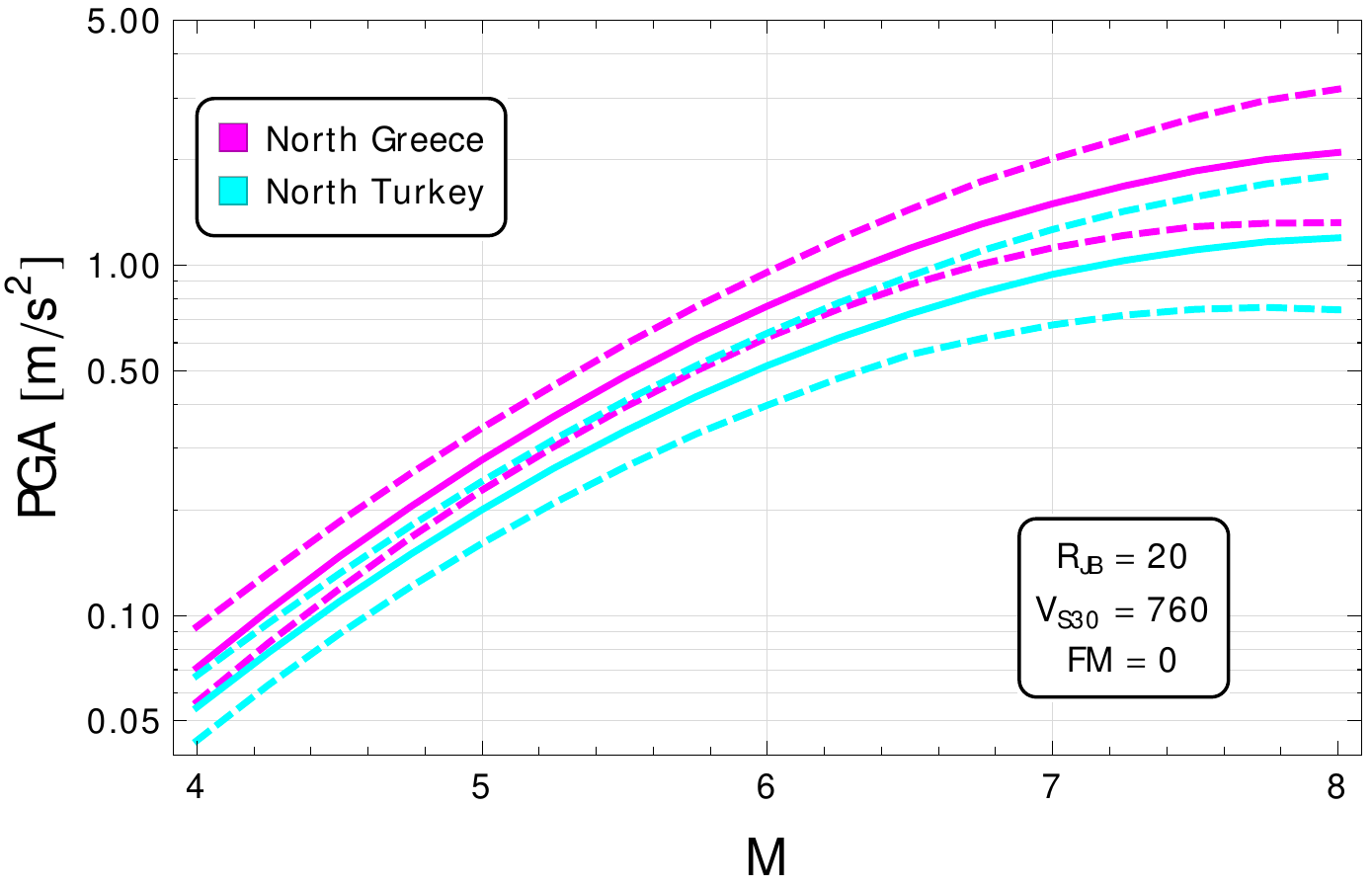}
  \end{center}
  \caption{Scaling of ground-motions predictions from the ``regional 3'' model. Solid lines are the median, dashed lines the 5\% and 95\% quantiles of the posterior predictive distribution.}
  \label{fig: scaling posterior}
\end{figure}

The parameters of the model in this work are estimated using Bayesian inference (e.g.\ \cite{Spiegelhalter2009}).
This means that all parameters of the model are associated with a (posterior) distribution, which is a measure of (epistemic) uncertainty associated with the parameters.
This uncertainty translates into uncertainty regarding the median predictions, and should not be neglected.
Figure \ref{fig: scaling posterior} shows the posterior distribution of the magnitude scaling for two regions, by plotting the the 5\%, 50\% and 95\% quantiles of the predicted PGA values based on the posterior distribution of the coefficients.
We only show two regions because it is difficult to make out differences in the posterior predictions if more than two scaling distributions are shown.
It is obvious from Figure \ref{fig: scaling posterior} that there is some overlap in the posterior distributions for Northern Greece and Northern Turkey, but this overlap is small.
This indicates that the median predictions for both regions are indeed different.
However, based on Figure \ref{fig: scaling posterior} alone, we can only make statements for the two regions for $R_{JB} = 20$km.

\begin{figure}
  \begin{center}
    \includegraphics[width = 0.95\textwidth]{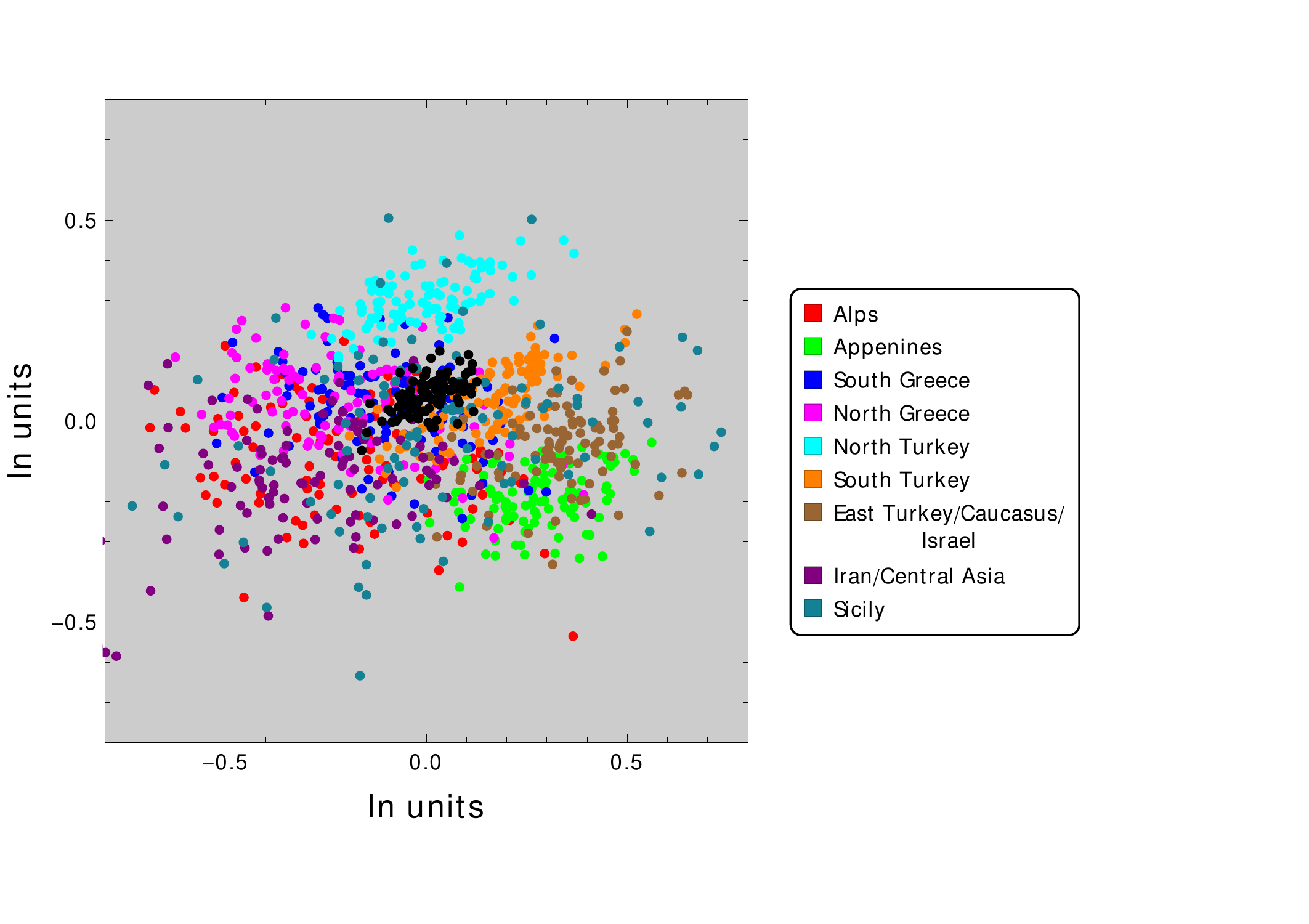}
  \end{center}
  \caption{Sammon's map of regionalized predictions, based on samples from the posterior distribution of the coefficients for each region, together with the global model (black). Distance between two points on the map correspond to the average difference in median predictions for two GMPEs (which differ in the values of the coefficients). The axes do not have a physical meaning, only relative differences are of interest. For details on GMPEs and Sammon’s maps, see \cite{Scherbaum2010}.}
  \label{fig: sammonsmap}
\end{figure}

To investigate the differences in predictions of the different regions over a broader magnitude/distance range, we look at a Sammon's map \cite{Sammon1969,Scherbaum2010}.
Therefore, for each region we draw 100 samples from the posterior distribution of the respective coefficients, and calculate predictions at magnitudes $M = 4.5,5.,\ldots,7.5$ and distances $R_{JB} = 5.,10.,20.,50.,75.,100.,150.,200.$.
For each sample, this gives a point in a 56-dimensional ground-motion space, which is projected into two dimensions by the Sammon's map algorithm -- in short, a two-dimensional projection is sought that preserves the differences between all points in the high-dimensional space (see \cite{Sammon1969,Scherbaum2010} for details).
In addition to the regional samples, we draw 100 samples from the posterior distribution of the global model.
The final map is shown in Figure \ref{fig: sammonsmap}.
This map is a representation of the mean difference between any two models over the selected magnitude/distance range.
The distance between two points on the map represents the average difference between their corresponding models over the whole magnitude/distance range in ln units.
As one can see, the predictions of some regions overlap quite a bit (e.g.\ North and South Greece), but many regions have distinct predictions, which are also relatively far away from the global model.
In particular, the Appenines and Northern Turkey occupy distinct sections on the map, indicating that their predictions are quite different from the other regions.
We can also see that the posterior distribution of median predictions for Sicily are relatively broadly distributed, which suggest a larger uncertainty.
This is a reflection of the small amount of data for Sicily.
Overall, the Sammon's map shows that the median predictions (both the average and the posterior distribution) are quite different for different regions, which will have effects on hazard calculations, since these average over a wide range of $M/R$-scenarios.
In addition, the regionalized model has a lower standard deviation than the global model, which also effects hazard calculations \cite{Bommer2006}.

We assessed the different models using approximations of predictive error (WAIC).
This measure estimates the error made for predictions on a new data set.
We believe that it is beneficial to look at measures such as these to assess the quality of a model and not only look at residuals and estimated standard deviations.
On the other hand, one should not rely on WAIC (or similar measures) alone, since they do not capture the extrapolation capability of a model.
This is demonstrated in the model with independent regional coefficients, which performs well in terms of WAIC, but does not extrapolate well to large magnitudes for Sicily (cf.\ Figure \ref{fig: comparison}).

\section*{Data and Resources}

Data comes from the Resorce database \cite{Akkar2014a}.
Figures are prepared using the program Mathematica (\url{https://www.wolfram.com/mathematica/}).
The values of $\Delta\widehat{elpd}_{WAIC}$ are calculated using the R-package \emph{loo}, version 0.1.3.
The Sammon's map is calculated using the R-package \emph{MASS}, version 7.3.
Inference is carried out using the program STAN, version 2.5.0 \cite{Team2015}.

\section*{Acknowledgements}

We would like to thank Norm Abrahamson, Christine Goulet and Justin Hollenback for discussions on the topic of regionally vaying GMPEs.
We would also like to thank the reviewers Sinan Akkar and John Douglas for their thoughtful comments, which greatly helped to improve the manuscript.

\end{document}